\documentclass[11pt,fleqn,a4paper]{article}

\usepackage{graphicx}
\usepackage{amsfonts}
\usepackage[mathscr]{euscript}
\usepackage{here}
\usepackage[numbers]{natbib}
\usepackage{notoccite}

\newcommand{\bb}[1]{\mbox{\boldmath $#1$\unboldmath}}

\newcommand{\dd}{\displaystyle}
\newcommand{\vp}{\vspace{4pt}}

\newcommand{\tr}{\makebox{tr}}

\renewcommand\appendix{\par
  \setcounter{section}{0}
  \setcounter{subsection}{0}
  \setcounter{figure}{0}
  \setcounter{table}{0}
  \renewcommand\thesection{Appendix \Alph{section}}
  \renewcommand\thefigure{\Alph{section}\arabic{figure}}
  \renewcommand\thetable{\Alph{section}\arabic{table}}
}

\title{Plasticity and non-Schmid effects}

\author{Stefan C. Soare\,\,\thanks{Technical Univ. of Cluj-Napoca, Romania,  stef$\_$soare@yahoo.com}}
\date{\small{30 June 2013}}

\begin{document}
\maketitle

\begin{abstract}
The structure of the overall response of a metal polycrystal is
analyzed under the assumption that activity on the slip systems of
its constituents is governed by a generalized Schmid law. In the
resulting macro-model of metal plasticity, the overall rate of
plastic deformation is related to the stress-gradient of the yield
surface via a generalized, but still associated flow rule. While
the overall rate of plastic deformation is no longer collinear with the
stress-gradient of the yield surface, its deviation is characterized by
an additional macro-variable, representative of the rate of
the non-Schmid effects taking place at constituent level.
\end{abstract}


\section{Introduction}

\noindent
The basic phenomenology of crystal plasticity was clarified in the early 1900's
when it was recognized that crystals deform by plastic shearing along crystallographic directions, e.g.,
\citet{EwingRosenhain}, \citet{Taylor38}, see also the brief but very
informative introduction into the early history of crystal plasticity in \citet{Asaro81}.
At about the same time it was proposed also, and later confirmed by direct observations,
that plastic shear is, in fact, the result of the motion of numerous dislocations
(one dimensional crystal defects) along crystallographic planes and directions,
a process called slip. The specific crystallographic planes and directions along which a crystal can be deformed
by plastic shearing are called slip systems. Schmid law is the constitutive link between
the crystal kinematics and the stress field generated within the crystal by external loading.
It states that shearing on a slip system can be initiated only
if its corresponding resolved shear stress reaches a critical value, \citet{Schmid1924}.
When all slip systems are considered, Schmid law defines a yielding (or activation) surface in stress space
and, in the rigid-plastic approximation, the rate of deformation lies within the normal cone at the current stress state on the yielding surface.
When aggregates of many crystals are considered, and \citet{Taylor38} homogenization principle is employed (each constituent experiences the deformation imposed at the boundary of the aggregate) this geometrical structure propagates virtually intact at macro-level:
the macro-rate of deformation is along the exterior normal to the overall yield surface, Bishop and Hill(1951a,b).

The theory was later extended to allow for arbitrary elastic strains, \citet{Rice71} and \citet{HillRice72}, and for general conditions
for estimating the overall response of an aggregate, \citet{Hill72}.
However, in the elastic-plastic context, the normality structure of the theory does not seem to follow as straightforwardly:
in \citet{Rice71} it was conditioned by the assumption that "the rate of progression of any local micro-structural rearrangement
within the material is dependent on the current stress state only through the thermodynamic force conjugate to the extent of that
rearrangement";
in \citet{HillRice72} it was recognized that normality depends on the evolution law assigned to the slip systems of a constituent.
The matter is reconsidered in \citet{Soare2013} where the structure of the
macroscopic constitutive model is deduced for a quite general class of slip system evolutions;
in particular, if the slip systems convect with the lattice of a constituent,
the macro-model enjoys the normality structure.

On the other hand, phenomena like cross-slip of dislocations, \citet{AsaroRice},
or pressure dependence of yielding, \citet{SR84}, imply that Schmid law may provide only an approximate description
of the forces governing slip activity; hence the normality rule also may provide just an approximate description of the direction of plastic flow.
The question is then: what is the precise relationship, if any, between the direction of plastic flow and the yield surface,
at both single crystal and macro levels when generalizations of Schmid activation criterion are considered ?
It appears that the question is not trivial since violations of Schmid law (or non-Schmid effects) at crystal level have been usually
interpreted as an indication that in continuum models of metal plasticity the yield surface
no longer determines the direction of plastic flow, via the normality rule, and that an additional constitutive element,
in the form of a plastic potential, would be required to characterize the direction of plastic flow, e.g., \citet{Vitek2004}, \citet{Groger2008}.

Here, we consider a polycrystal aggregate for which a generalized Schmid law controls slip at constituent level,
and aim at deducing the structure of the overall response of the aggregate.
The approach is similar to that in \citet{Soare2013} but the framework is different.
Since the symmetry of the instantaneous elastic moduli is crucial for both theoretical and practical considerations,
and with an eye to further applications, the constitutive response of a constituent will be explicitly hyper-elastic.
A general theory of the elastic-plastic response of such aggregates has been described by \citet{HillRice73}.
Most of the present developments will be featured in the context of this theory.
At its foundations lies the parametrization of the plastic state;
while in the cited work this is left at an abstract level, the aim being at a general theory,
here the multiplicative elastic-plastic decomposition of \citet{Rice71} is employed explicitly to construct
parameterizations of the plastic state at crystal level.

\section{Single crystal model: basic equations}

\noindent We begin our analysis from what is actually observed.
Thus we write the basic equations characterizing the response of a
crystal by first referring them to a spatial (or laboratory) frame
(eulerian description). Then considering a crystal deformed
plastically by shearing along crystallographic directions, an
observer of the resulting motion writes, \citet{HillRice72},
\citet{AsaroRice}, \citet{Asaro}:
\begin{equation}
\bb{d} = \bb{d}^{e}+\widehat{\bb{d}}^{p},\,\,\, \bb{w} = \bb{w}^e+\bb{w}^p,\,\,\,
\bb{l}^p:=\widehat{\bb{d}}^p +\bb{w}^p
\label{Decomp}
\end{equation}
\begin{equation}
\widehat{\bb{d}}^{p} := \sum_{\alpha\in \mathcal{A}} \dot{\gamma}^{\alpha}\bb{a}^{\alpha},\,\,\, \bb{w}^{p} := \sum_{\alpha\in \mathcal{A}} \dot{\gamma}^{\alpha}\bb{b}^{\alpha}
\label{Flow}
\end{equation}
\begin{equation}
\bb{\dot{\tau}}^{Je}:=
\bb{\dot{\tau}}+\bb{\tau}\bb{w}^e-\bb{w}^e\bb{\tau} = \bb{K}^J:\left(\bb{d}-\widehat{\bb{d}}^{p}\right)
\label{Elast}
\end{equation}
where: $\bb{l}=\bb{d}+\bb{w}$ represents the decomposition of the spatial gradient
$\bb{l}$ of the local velocity field into rate of deformation and spin;
$\mathcal{A}$  denotes the set of active slip systems at a material point of the crystal,
$\dot{\gamma}^{\alpha}$ denotes the rate of plastic shear on slip system $\alpha$, and
\begin{equation}
\bb{a}^{\alpha}: = \left(\bb{m}^{\alpha}\otimes\bb{n}^{\alpha}+\bb{n}^{\alpha}\otimes\bb{m}^{\alpha}\right)/2,\,\,\,
\bb{b}^{\alpha}: = \left(\bb{m}^{\alpha}\otimes\bb{n}^{\alpha}-\bb{n}^{\alpha}\otimes\bb{m}^{\alpha}\right)/2.
\label{aalphaspatial}
\end{equation}
$\left\{\bb{m}^{\alpha},\bb{n}^{\alpha}\right\}$ and
$\left\{-\bb{m}^{\alpha},\bb{n}^{\alpha}\right\}$ are viewed as
two distinct slip systems; hence $\dot{\gamma}^{\alpha}> 0$, for
all $\alpha\in\mathcal{A}$.

$\bb{K}^J$ is the tensor of instantaneous elastic moduli associated with the Jaumann rate of the Kirchhoff stress $\bb{\tau}:=J\bb{\sigma}$,
where $J$ is the determinant of the deformation gradient (with respect to some reference configuration)
and $\bb{\sigma}$ is the Cauchy stress.
Eq.(\ref{Elast}) reproduces the point of view of an observer attached to a frame spinning at the rate $\bb{w}^e$,
the spin of the crystal lattice with respect to the material.
The point of view of an observer attached to a frame spinning with the material is obtained by
rearranging eq.(\ref{Elast}) in the form
\begin{equation}
\bb{\dot{\tau}}^J:=\bb{\dot{\tau}}+\bb{\tau}\bb{w}-\bb{w}\bb{\tau} = \bb{K}^J:\left(\bb{d}-\bb{d}^p\right)
\label{MaterialDP}
\end{equation}
where it has been defined
\begin{equation}
\bb{d}^p:=\widehat{\bb{d}^p} -\left(\bb{K}^J\right)^{-1}:\left(\bb{\tau}\bb{w}^p-\bb{w}^p\bb{\tau}\right)
\label{MaterialDP2}
\end{equation}
Of further interest is the restatement of the above stress-strain relationship in terms of the Truesdell stress rate.
A rearrangement of eq.(\ref{MaterialDP}) yields
\begin{equation}
\bb{\dot{\tau}}^L:=\bb{\dot{\tau}}-\bb{l}\bb{\tau}-\bb{\tau}\bb{l}^T=\bb{K}^L:(\bb{d}-\bb{d}^{pL})
\label{TruesdellSS}
\end{equation}
where, based on the relationship $\bb{\dot{\tau}}^L = \bb{\dot{\tau}}^J-\left(\bb{\tau}\bb{d}+\bb{d}\bb{\tau}\right)$,
for any symmetric second order tensor $\bb{d}^*$ the tensor of instantaneous moduli $\bb{K}^L$ is defined by
$\bb{K}^L:\bb{d}^*:=\bb{K}^J:\bb{d}^*-\left(\bb{\tau}\bb{d}^*+\bb{d}^*\bb{\tau}\right)$.
The corresponding rate of plastic deformation, as measured by an observer attached to a Truesdell frame (i.e., a frame convecting with the motion), is
\begin{equation}
\bb{d}^{pL}:=\left(\bb{K}^L\right)^{-1}\bb{K}^J:\bb{d}^p = \widehat{\bb{d}}^{p}+\left(\bb{K}^L\right)^{-1}:\left(\bb{\tau}\bb{l}^{pT}+\bb{l}^p\bb{\tau}\right)
\label{TruesdellDP}
\end{equation}

\section{Single crystal model with elastic potential}
\label{S3}
\noindent
Defining an elastic potential requires a parametrization of the plastic state.
It seems that the most convenient way to achieve this is by employing the multiplicative elastic-plastic
decomposition of the deformation gradient with respect to a reference configuration.

Then let $\mathcal{B}_0\subset\mathbb{R}^3$ denote a reference configuration of a crystal subject to the motion
$x = \hat{x}(X,t)$, $ X\in\mathcal{B}_0$, $t\in[0,\infty)$.
The motion, $\mathcal{B}_0$ and the current configuration $\mathcal{B}_t:=\left\{x(X,t)|X\in\mathcal{B}_0\right\}$ of the crystal are referred to a common Cartesian (fixed) frame.
At each material particle $X$ a multiplicative decomposition of the deformation gradient $\bb{F}(X)$ is assumed to exist, \citet{Rice71}:
\begin{equation}
\frac{\partial \hat{x}}{\partial X}(X,t)=:\bb{F}(X,t) = \bb{F}^*(X,t)\bb{F}^p(X,t)
\label{MultDec}
\end{equation}
Thinking of the crystal as of a composite consisting of "material" and "lattice",
$\bb{F}^p$ is to represent the plastic shearing of material (through the crystal lattice)
and $\bb{F}^*$ the accompanying elastic distortion during which the lattice and material deform as one,
see \citet{AsaroRice} or \citet{Asaro} for a pictorial representation.
$\bb{F}^*$ incorporates also the local rotation of the crystal due to the constraints imposed by the neighboring material
(through boundary conditions).
In general, the decomposition (\ref{MultDec}) is not expected to be compatible,
in the sense that, in general, $\bb{F}^*$ and $\bb{F}^p$ may not be the gradients of some smooths
motions. If $T_X$ and $T_x$ denote, respectively, the tangent spaces at the reference and current configurations at particle $X$
and its position $x=\hat{x}(X,t)$, then the collection $\mathcal{B}^*:=\left.\left\{\bb{F}^*:T_X\longrightarrow T_x\,\right| X\in\mathcal{B}_0\right\}$
is usually referred to as \emph{the intermediate configuration} of the crystal.
Here the term "configuration" is employed loosely, since $\mathcal{B}^*$
is not a (differentiable) manifold in the Euclidean ambient space of the motion $\hat{x}$, e.g., \citet{AcharyaBassani}, \citet{Gupta2007}.

Taking the time rate in eq.(\ref{MultDec}), with $\bb{l}:=\bb{\dot{F}}\bb{F}^{-1}$, $\bb{l}^*:=\bb{\dot{F}}^*\bb{F}^{*-1}$,
$\bb{L}^P:=\bb{\dot{F}}^p\bb{F}^{p-1}$, one obtains
\begin{equation}
\bb{l} = \bb{l}^* + \bb{F}^*\bb{L}^p\bb{F}^{*-1}
\label{MultDecAdd}
\end{equation}
In particular, if the reference configuration is the current configuration,
then the above additive decomposition of $\bb{l}$ must be identical
with the decomposition $\bb{l}=\bb{l}^e+\bb{l}^p$
featured in eqs.(\ref{Decomp}) and (\ref{Flow}), since the two decompositions are supposed to describe the same kinematics.
Furthermore, this must hold for any reference configuration and so
$\bb{F}^*\bb{L}^p\bb{F}^{*-1} = \bb{l}^p = \sum_{\alpha\in\mathcal{A}}\dot{\gamma}^{\alpha}\bb{m}^{\alpha}\otimes\bb{n}^{\alpha}$,
which is equivalent to $\bb{L}^P = \sum_{\alpha\in\mathcal{A}}\dot{\gamma}^{\alpha}\left(\bb{F}^{*-1}:\bb{m}^{\alpha}\right)
\otimes\left(\bb{F}^{*T}:\bb{n}^{\alpha}\right)$.
Then given $\bb{m}^{0\alpha}$ and $\bb{n}^{0\alpha}$ unit vectors
in the slip and normal directions in the reference configuration $\mathcal{B}_0$,
the following (natural) evolution law is adopted for the slip systems of the crystal:
\begin{equation}
\begin{array}{l}
\dd\bb{m}^{\alpha} = \bb{F}^*:\bb{m}^{0\alpha},\,\,
\bb{n}^{\alpha} = \bb{F}^{*-T}:\bb{n}^{0\alpha}\,\Longrightarrow\,
\bb{L}^p = \sum_{\alpha\in\mathcal{A}}\dot{\gamma}^{\alpha}\bb{G}^{\alpha},\vp\\
\dd\makebox{where}\,\,\,\,\,
\bb{G}^{\alpha}:=\bb{m}^{0\alpha}\otimes\bb{n}^{0\alpha}
\end{array}
\label{SlipEvolv}
\end{equation}
Thus $\{\bb{m}^{\alpha},\bb{n}^{\alpha}\}$ convect with the lattice and an observer placed in a frame convecting
with the lattice will measure plastic strain as if this would take place in an undistorted (reference) crystal.

There are many other plausible evolution laws which could be assigned to the slip systems, \citet{HillRice72},
\citet{AsaroRice}. In the context of decomposition (\ref{MultDecAdd}), once such an evolution law
is assigned to the slip systems, then the corresponding characterization of $\bb{L}^p$ follows.
For example, if the slip systems are assumed to just rotate rigidly at the spin of the lattice,
a case investigated in \citet{Soare2013},
i.e., $\bb{\dot{m}}^{\alpha} = \bb{w}^*:\bb{m}^{\alpha}$ and $\bb{\dot{n}}^{\alpha} = \bb{w}^*:\bb{n}^{\alpha}$,
with $\bb{w}^*:=(\bb{l}^*-\bb{l}^{*T})/2$, then one can show that
$\bb{L}^p = \sum_{\alpha\in\mathcal{A}}\dot{\gamma}^{\alpha}\bb{U}^{*-1}\bb{R}^*_U\bb{G}^{\alpha}\bb{R}^{*T}_U\bb{U}^*$,
where $\bb{F}^*=\bb{R}^*\bb{U}^*$ is the polar decomposition of $\bb{F}^*$ and $\bb{R}^*_U$ is the rotation at spin
$(\bb{\dot{U}}^*\bb{U}^{*-1}-\bb{U}^{*-1}\bb{\dot{U}}^*)/2$.
\emph{Throughout the rest of this work it will be assumed
that the slip systems evolve according to eq}.(\ref{SlipEvolv}).
The main consequence of this assumption is that, in the absence of non-Schmid effects,
the structure of the constitutive system at micro and macro-levels enjoys the normality structure, \citet{HillRice72}, \citet{Soare2013}.

As stress and strain measures on $\mathcal{B}_0$ we adopt the conjugate pair of the symmetric Kirchhoff stress and Green strain,
that is, $\bb{S}:=\bb{F}^{-1}(J\bb{\sigma})\bb{F}^{-T}$, $\bb{E}:=(\bb{F}^T\bb{F}-\bb{I})/2$.
Although the next arguments extend to other stress-strain measures, \citet{Hill68a}, formulas and derivations are
simplest when the pair $\{\bb{S},\bb{E}\}$ is employed. The reason for this is that simple and explicit pull-back/push-forward formulas relate
$\{\bb{S},\bb{\dot{E}}\}$ to $\{\bb{\sigma},\bb{d}\}$, for any reference configuration, where the fundamental formula
$\bb{D}:=\bb{\dot{E}} = \bb{F}^T\bb{d}\bb{F}$ is recalled.
The same pull-back formula must hold for the rate $\widehat{\bb{d}}^p$ defined in eqs.(\ref{Decomp}) and (\ref{Flow}) and hence
its image in the reference configuration is
\begin{equation}
\begin{array}{l}
\dd\widehat{\bb{D}}^p := \bb{F}^T\,\widehat{\bb{d}}^p\,\bb{F} = \sum_{\alpha\in\mathcal{A}}\dot{\gamma}^{\alpha}\bb{A}^{\alpha},\vp\\
\dd\makebox{with}\,\,\,\,\,
\bb{A}^{\alpha}:=\bb{F}^T\bb{a}^{\alpha}\bb{F} = \bb{F}^{pT}\frac{1}{2}\left(\bb{C}^*\bb{G}^{\alpha}+\bb{G}^{\alpha T}\bb{C}^*\right)\bb{F}^p
\end{array}
\label{HatDP}
\end{equation}
where $\bb{C}^*:=\bb{F}^{*T}\bb{F}^*$. With $\bb{K}^L$ now \emph{defined} by
$\bb{K}^L:\bb{a}$ $=$ $\bb{F}$ $\left[\bb{K}:\left(\bb{F}^T\bb{a}\bb{F}\right)\right]\bb{F}^T$,
for any second order symmetric tensor $\bb{a}$, $\widehat{\bb{D}}^{pL} := \bb{F}^{T}\bb{d}^{pL}\bb{F}$
then acquires the following representation
\begin{equation}
\begin{array}{lll}
\dd\widehat{\bb{D}}^{pL} &=& \widehat{\bb{D}}^{p}+\bb{K}^{-1}:\left(\bb{S}\bb{F}^{pT}\bb{L}^{pT}\bb{F}^{p-T}+
\bb{F}^{p-1}\bb{L}^p\bb{F}^p\bb{S}\right) =\vp\\
\dd&=&\sum_{\alpha\in\mathcal{A}}\dot{\gamma}^{\alpha}
\left[\bb{A}^{\alpha}+\bb{K}^{-1}:\left(\bb{S}\bb{H}^{\alpha T}+\bb{H}^{\alpha}\bb{S}\right)\right]
\end{array}
\label{DPrefTruesdell2}
\end{equation}
obtained by employing eqs.(\ref{TruesdellDP}) and (\ref{SlipEvolv}), and where
\begin{equation}
\bb{H}^{\alpha}:=\bb{F}^{p-1}\bb{G}^{\alpha}\bb{F}^p
\label{Halpha}
\end{equation}

The yet unspecified Lagrangian tensor of elasticity $\bb{K}$ is now defined by assuming
the crystal enjoys a hyper-elastic response. The safest approach to defining the latter
is via the theory of elastic-plastic deformation developed by \citet{HillRice73}.
Here, as a particularization of this theory, it is assumed that there exists a symmetric second order tensor $\bb{E}^p$,
which is supposed to characterize the current plastic state of the crystal at particle $X$,
and that there exists an elastic potential $\phi=\phi\left(\bb{E},\bb{E}^p\right)$,
such that at any moment during the motion the stress state at particle $X$ is given by
\begin{equation}
\bb{S}=\frac{\partial\phi}{\partial\bb{E}}\left(\bb{E},\bb{E}^p\right)\,\,\,\Longrightarrow\,\,\,
\bb{\dot{S}} = \bb{K}:\left[\bb{\dot{E}}-\bb{K}^{-1}\bb{K}^p:\bb{\dot{E}}^p\right] =
\bb{K}:\left[\bb{D}-\bb{D}^{pL}\right]
\label{CrystalRateForm}
\end{equation}
where the implication follows by taking the time rate of the stress-strain relationship in potential form,
and where the following definitions have been employed:
\begin{equation}
\bb{K}:=\frac{\partial^2\phi}{\partial\bb{E}\partial\bb{E}},\,\,\,\,
\bb{K}^p:=-\frac{\partial^2\phi}{\partial\bb{E}\partial\bb{E}^p},\,\,\,\,
\bb{D}^{pL}:= \bb{K}^{-1}\bb{K}^p:\bb{\dot{E}}^{p}
\label{CrystalRateForm2}
\end{equation}
In general, the equality $\bb{D}^{pL} = \bb{\dot{E}}^p$ does not hold,
unless $\bb{K}^p =\bb{K}$.
A sufficient condition for the latter equality to hold is that the elastic potential
be of the form $\phi(\bb{E},\bb{E}^p) := \phi^e(\bb{E}-\bb{E}^p)+\phi^p(\bb{E}^p)$.
Under the assumption, acceptable at crystal level, that plastic flow does not alter (significantly) the lattice of the crystal and hence its elasticity, throughout the rest of this work it will be assumed that \emph{the elastic potential is independent of the plastic state}, that is:
$\phi(\bb{E},\bb{E}^p) := \phi^e(\bb{E}^e)$, where $\bb{E}^e:=\bb{E}-\bb{E}^p$.
Thus, here, $\bb{E}^p$ can be regarded as a measure of plastic strain and $\bb{E}^e$ as a measure of the elastic strain.
It is underlined that this intuitive description, in terms of elastic and plastic strains,
is not valid in case the structure of the crystal is affected significantly by plastic flow.

Regarding the plastic parameter $\bb{E}^p$, this will be \emph{defined},
with respect to the chosen reference configuration $\mathcal{B}_0$,
as the solution of the Cauchy problem
\begin{equation}
\bb{\dot{E}}^{p}(X,t) =
\widehat{\bb{D}}^{pL},\,\,\,\,\bb{\bb{E}}^p(X,t_0) = \bb{0}
\label{PlasticStrain}
\end{equation}
Once evolution laws have been assigned to the shear rates $\dot{\gamma}^{\alpha}$, the above problem
characterizes uniquely, via eq.(\ref{DPrefTruesdell2}), the current value of the plastic parameter $\bb{E}^p$.
This parametrization has the important property that the rate form of the stress-strain relationship in eq.(\ref{CrystalRateForm})
is a perfect image in the reference configuration of the relationship in eq.(\ref{TruesdellSS}),
since $\bb{D}^{pL} = \bb{\dot{E}}^p = \widehat{\bb{D}}^{pL}$,
being thus compatible with the kinematics described in the previous section.
On the other hand, one can imagine many other definitions for $\bb{E}^p$,
each being associated with a corresponding elastic-plastic additive decomposition,
and one further example, popular in many investigations based on crystal plasticity, is described in the Appendix to this work.

\section{Slip activation criterion:\\
Incorporating non-Schmid effects}
\label{SchmidLaw}

\noindent
Schmid law states that slip system $\{\bb{m}^{\alpha},\bb{n}^{\alpha}\}$ is inactive as long as $\tau^{\alpha}<\tau^{\alpha}_{cr}$,
where $\tau^{\alpha}:=\bb{\sigma}\cdot\left(\bb{m}^{\alpha}\otimes\bb{n}^{\alpha}\right) = \bb{\sigma}\cdot\bb{a}^{\alpha}$
is the resolved shear stress on slip system $\alpha$; once $\tau^{\alpha}$ reaches the critical value $\tau^{\alpha}_{cr}$,
slip system $\alpha$ becomes (potentially) active.
This activation criterion is accurate as long the motion of the dislocations associated with slip system $\alpha$ is restricted
to the slip plane.
On the other hand, the motion of dislocations may exhibit deviations from this course.
For example, based on atomistic simulations of the motion of $1/2\hspace{-4pt}<\hspace{-4pt}111\hspace{-4pt}>$ screw dislocations,
\citet{Vitek2004} and \citet{Groger2008} describe a more complex phenomenology of yielding in b.c.c. lattices,
where the non-planar spreading of the dislocation cores leads to violations of Schmid'law, see also the recent review in \citet{Bassani2011}.
More precisely, besides $\tau^{\alpha}$, other shear components of the stress state are shown to influence the motion of dislocations
and these can lead to the switching of the slip plane (cross-slip). To model these phenomena at crystal level,
the cited works employ particular forms of an activation criterion proposed in \citet{Qin}, a work, in turn,
inspired by a continuum theory of cross-slip effects developed in \citet{AsaroRice}.
\citet{Qin} extend Schmid law in the form
\begin{equation}
\hspace{-1cm}
\bb{\sigma}\cdot \left(\bb{m}^{\alpha}\otimes\bb{n}^{\alpha}+\sum_{q}a_q^{\alpha}\bb{m}^{\alpha}_q\otimes\bb{n}^{\alpha}_q\right) \left\{
\begin{array}{l}
< \tau_c^{\alpha}, \,\,\,\makebox{slip system is inactive}\\
= \tau_c^{\alpha},\,\,\,\makebox{slip system is (potentially) active}
\end{array}\right.
\label{GeneralSchmid0}
\end{equation}
where $\{\bb{m}^{\alpha}_q,\bb{n}^{\alpha}_q\}$ are vectors that sample those components of the stress state
which may affect the yielding of the crystal, and $a_q^{\alpha}$'s are scalar constitutive functions.
More generally, but retaining the resolved shear stress as the main driving force, the criterion for slip system
$\{\bb{n}^{\alpha},\bb{m}^{\alpha}\}$ can be restated as
\begin{equation}
\hspace{-1cm}
\bb{\sigma}\cdot \bb{a}^{\alpha} +\zeta^{\alpha}(\bb{\sigma},\bb{p})\left\{
\begin{array}{l}
< \tau_c^{\alpha}\left[1+\zeta^{\alpha}_h(\bb{\sigma},\bb{p}_h)\right], \,\,\,\makebox{slip system is inactive}\\
= \tau_c^{\alpha}\left[1+\zeta^{\alpha}_h(\bb{\sigma},\bb{p}_h)\right],\,\,\,\makebox{slip system is (potentially) active}
\end{array}\right.
\label{GeneralSchmid}
\end{equation}
where the arguments $\bb{p}$ and $\bb{p}_h$ incorporate all the
additional structural parameters that $\zeta^{\alpha}$ and
$\zeta^{\alpha}_h$ may depend upon, e.g., the
$\{\bb{m}^{\alpha}_q,\bb{n}^{\alpha}_q\}$ vectors. The functions
$\zeta^{\alpha}$ represent the contribution of the stress
components affecting the configuration of the dislocation cores,
whereas $\zeta^{\alpha}_h$ represent deviations that may be
related to the hardening state of the crystal. For example, for
$\zeta^{\alpha}\equiv0$ and
$\zeta^{\alpha}_h(\bb{\sigma},\bb{p}_h) =
\zeta(\bb{\sigma}):=-q\,\tr(\bb{\sigma})$, with $q>0$ a
\emph{constant} material parameter, one recovers an extension of
Schmid's criterion considered in \citet{SoareBarlat2013} and shown
to reproduce at macro-level the criterion of \citet{SR84}.

By employing the pull-back of $\bb{a}^{\alpha}$ in eq.(\ref{HatDP}), so that
$\tau^{\alpha} = \bb{\sigma}\cdot\bb{a}^{\alpha} = (1/J)\bb{S}\cdot\bb{A}^{\alpha}$, the extension
in eq.(\ref{GeneralSchmid}) is reformulated in the reference configuration as follows
\begin{equation}
\bb{S}\cdot \bb{A}^{\alpha} + \hat{\zeta}^{\alpha}(\bb{S},\bb{P})\leq \tau_{cr}^{\alpha}\left[1+\hat{\zeta}^{\alpha}_h(\bb{S},\bb{P}_h)\right]
\label{GeneralSchmid2}
\end{equation}
where $\hat{\zeta}^{\alpha}(\bb{S},\bb{P}):=\zeta^{\alpha}(\bb{F}\bb{S}\bb{F}^T, \bb{F}\bb{p}\bb{F}^T)$, etc.
It is assumed the $\zeta^{\alpha}$'s and the $\zeta^{\alpha}_h$'s are objective functions (like those in eq.(\ref{GeneralSchmid0})),
so that the functions $\hat{\zeta}^{\alpha}$ and $\hat{\zeta}^{\alpha}_h$ are well defined (invariant to any superimposed rotation).

The above extension will be more convenient for the next derivations, since it does not depend explicitly on the volumetric
part $J$.
However, strictly speaking, if $\zeta^{\alpha}\equiv 0$ and $\zeta^{\alpha}_h\equiv 0$ are to yield the classical Schmid law,
the two activation criteria in eqs.(\ref{GeneralSchmid}) and (\ref{GeneralSchmid2}) are slightly different.
Indeed, in this case, eq.(\ref{GeneralSchmid2}) reads in the current configuration: $\bb{\sigma}\cdot \bb{a}^{\alpha}\leq \tau^{\alpha}_{cr}/J$.
Thus the criterion in eq.(\ref{GeneralSchmid2}) features an additional pressure dependence by comparison with the classical Schmid criterion.
Arguments for substituting the Cauchy stress with the Kirchhoff stress in the activation criterion
are given in \citet{Soare2013}; these show that eq.(\ref{GeneralSchmid2}) is a valid extension of Schmid law
when other non-Schmid effects, represented by the functions $\zeta^{\alpha}$ and $\zeta^{\alpha}_h$, are neglected.

Finally, considering the thermodynamic consistency of the present model, let us note that,
by eqs.(\ref{DPrefTruesdell2}) and (\ref{GeneralSchmid2}), during plastic flow there holds
\begin{equation}
\bb{S}\cdot\widehat{\bb{D}}^{pL} =
\sum_{\alpha\in\mathcal{A}}\dot{\gamma}^{\alpha}
\left[\tau^{\alpha}
+\left(\bb{K}^{-1}:\bb{S}\right)\cdot\left(\bb{S}\bb{H}^{\alpha T}+\bb{H}^{\alpha}\bb{S}\right)\right]
\label{WorkRate}
\end{equation}
In general, for the stress levels sustained by metals,
the second term between the square brackets is much smaller than the first, see also Appendix for its analysis;
it can be concluded that $\bb{S}\cdot\widehat{\bb{D}}^{pL}>0$.
Let us note also that in the present framework the resolved shear stress $\tau^{\alpha}$ can be interpreted as a
thermodynamic force conjugate to the corresponding shear displacement $\gamma^{\alpha}$ only modulo an elastic slip system distortion (the second term between the square brackets). This relaxes the stronger assumption in \citet{Rice71} that $\tau^{\alpha}$ be precisely conjugate to $\gamma^{\alpha}$.

\section{Flow/Normality rule}
\label{CrystalFlowRule}

\noindent
Next, the relationship between the direction of plastic flow and the yielding (or activation) surface of the crystal
is investigated.
Following a convention in \citet{HillRice73}, see also \citet{Soare2013}, the symbol $\delta$ will be employed in this work to denote the
$\theta$-time rate of an object along an elastic deformation history at its starting point, $\theta=0$,
while keeping the reference particle fixed. Thus
\begin{equation}
\delta\bb{S}:=\lim_{\theta\rightarrow 0}\frac{1}{\theta}\left[\bb{S}^{\theta}-\bb{S}(t)\right]
\end{equation}
signifies an elastic (stress) direction if $\bb{S}^{\theta} := \bb{S}(t+\theta)$, $\theta>0$, is an elastic stress trajectory for at least some
interval $(0,\theta_M]\ni\theta$, with $\theta_M>0$.
Then taking the $\theta$-time rate of the elastic stress-strain
relationship in eq.(\ref{CrystalRateForm}), in potential form, obtains
\begin{equation}
\delta\bb{S} = \bb{K}:\delta\bb{E} = (1/2)\bb{K}:\delta\bb{C}
\label{DeltaS}
\end{equation}
Conversely, for any elastic direction $\delta\bb{S}$ there exists an elastic trajectory $\bb{S}^{\theta}$
having $\delta\bb{S}$ as its rate at $\theta=0$.

Let $\bb{S}:=\bb{S}(t)$ represent the current stress; assuming that at $t$ the crystal is in a state of plastic deformation,
consider next an arbitrary elastic trajectory $\bb{S}^{\theta}=\bb{S}(t+\theta)$, $\theta>0$, originating
at the current stress $\bb{S}$. By eq.(\ref{GeneralSchmid2}), there holds:
\begin{displaymath}
\frac{\bb{S}^{\theta}\cdot\left(\bb{A}^{\alpha}\right)^{\theta}+\hat{\zeta}^{\alpha}\left(\bb{S}^{\theta},\bb{P}^{\theta}\right)}
{1+\hat{\zeta}^{\alpha}_h\left(\bb{S}^{\theta},\bb{P}^{\theta}_h\right)}<\tau^{\alpha}_{cr} =
\frac{\bb{S}\cdot\bb{A}^{\alpha}+\hat{\zeta}^{\alpha}\left(\bb{S},\bb{P}\right)}
{1+\hat{\zeta}^{\alpha}_h\left(\bb{S},\bb{P}_h\right)}
\end{displaymath}
whence by dividing by $\theta$ and taking the limit $\theta\longrightarrow 0$:
\begin{equation}
\begin{array}{l}
\dd\delta\left[\frac{\bb{S}\cdot\bb{A}^{\alpha}+\hat{\zeta}^{\alpha}\left(\bb{S},\bb{P}\right)}
{1+\hat{\zeta}^{\alpha}_h\left(\bb{S},\bb{P}_h\right)}\right]\leq 0\,\,\,\iff\vp\\
\dd\,\,\,\,\,\iff\,\,\,\,\delta\left[\bb{S}\cdot\bb{A}^{\alpha}+\hat{\zeta}^{\alpha}\left(\bb{S},\bb{P}\right)\right]-
\tau^{\alpha}_{cr}\,\delta\left[\hat{\zeta}^{\alpha}_h\left(\bb{S},\bb{P}_h\right)\right]\leq 0
\end{array}
\label{DeltaInequality}
\end{equation}
With $\bb{C}=\bb{F}^T\bb{F} = \bb{F}^{pT}\bb{C}^*\bb{F}^p$ there follows $\delta\bb{C}^* = \bb{F}^{p-T}\delta\bb{C}\bb{F}^{p-1}$
and then, by employing eq.(\ref{DeltaS}):
\begin{equation}
\delta\bb{A}^{\alpha} = (1/2)\left(\delta\bb{C}\bb{H}^{\alpha}+\bb{H}^{\alpha T}\delta\bb{C}\right) =
\left(\bb{K}^{-1}:\delta\bb{S}\right)\bb{H}^{\alpha}+\bb{H}^{\alpha T}\left(\bb{K}^{-1}:\delta\bb{S}\right)
\label{DeltaA}
\end{equation}
where, recall, $\bb{H}^{\alpha}$ was defined in eq.(\ref{Halpha}).
Since the parameters assembled in $\bb{P}$ are in general tensor products among lattice vectors, like in eq.(\ref{GeneralSchmid0}),
then similarly to the above formula for $\bb{A}^{\alpha}$ one can assume without loss of generality that
there exist tensors $\bb{B}$ and $\bb{B}_h$ of adequate dimensions such that $\delta\bb{P} = \bb{B}:\delta\bb{S}$
and $\delta\bb{P}_h = \bb{B}_h:\delta\bb{S}$. Then one can write
\begin{equation}
\delta\left[\hat{\zeta}^{\alpha}\left(\bb{S},\bb{P}\right)\right] = \bb{\Pi}^{\alpha}\cdot\delta\bb{S},\,\,\,\makebox{where}\,\,\,\,
\bb{\Pi}^{\alpha}:=\frac{\partial\hat{\zeta}^{\alpha}}{\partial\bb{S}}+\bb{B}^T:\frac{\partial\hat{\zeta}^{\alpha}}{\partial\bb{P}}
\label{DeltaZeta}
\end{equation}
\begin{equation}
\delta\left[\hat{\zeta}^{\alpha}_h\left(\bb{S},\bb{P}_h\right)\right] = \bb{\Pi}^{\alpha}_h\cdot\delta\bb{S},\,\,\,\makebox{where}\,\,\,\,
\bb{\Pi}^{\alpha}_h:=\frac{\partial\hat{\zeta}^{\alpha}_h}{\partial\bb{S}}+\bb{B}^T_h:\frac{\partial\hat{\zeta}^{\alpha}_h}{\partial\bb{P}_h}
\label{DeltaZetaH}
\end{equation}
Substituting eqs.(\ref{DeltaA}-\ref{DeltaZetaH}) into eq.(\ref{DeltaInequality}), it may be deduced, by the symmetry of $\bb{K}^{-1}$,
that for every active slip system there holds
\begin{displaymath}
\delta\bb{S}\cdot\left(\bb{A}^{\alpha}+\bb{K}^{-1}:\widehat{\bb{H}}^{\alpha}+\bb{\Pi}^{\alpha}+\tau^{\alpha}_{cr}\,\bb{\Pi}^{\alpha}_h\right)\leq 0,\,\,\,\,
\makebox{with}\,\,\,\widehat{\bb{H}}^{\alpha}:=\bb{S}\bb{H}^{\alpha T}+\bb{H}^{\alpha}\bb{S}
\end{displaymath}
Multiplying, for each $\alpha\in\mathcal{A}$, the above inequality with the corresponding shear rate $\dot{\gamma}^{\alpha}$
and then summing over the set of active slip systems we finally deduce, by employing eq.(\ref{DPrefTruesdell2}):
\begin{equation}
\delta\bb{S}\cdot\left(\widehat{\bb{D}}^{pL}+\bb{Z}\right)\leq 0,\,\,\,\,\,\makebox{where}\,\,\,\,\,
\bb{Z}:=\sum_{\alpha\in\mathcal{A}}\dot{\gamma}^{\alpha}\left(\bb{\Pi}^{\alpha}+\tau^{\alpha}_{cr}\,\bb{\Pi}^{\alpha}_h\right)
\label{NormalityRef}
\end{equation}
The above inequality is valid for arbitrary elastic directions $\delta\bb{S}$ issued at the current yielding stress $\bb{S}$.
As such, it is equivalent to a flow rule:
The bracketed object lies within the normal cone to the activation surface of the crystal, at the current stress.
One may note also that the elastic variation $\delta\bb{A}^{\alpha}$ depends essentially
on the evolution law assigned to the slip systems of the crystal.
As such, under the assumptions in eqs.(\ref{SlipEvolv}) and (\ref{PlasticStrain}), inequality (\ref{NormalityRef}) is optimal:
the only source for deviations from the classical normality rule are the non-Schmid effects.
Other forms of slip system evolution, or other parameterizations of the plastic state lead, in general, to additional deviations
from the normal cone to the activation surface, \citet{Soare2013} and Appendix.

\section{Overall characteristics:\\
Elastic potential and flow rule}

\noindent
The next goal is to deduce the \emph{essential}, in the sense of \citet{Hill67}, structure of the overall response of a polycrystal
whose constituents feature the just described constitutive response.
This will be done here by estimating the response of the polycrystal when subject to uniform (or homogeneous) displacement boundary conditions.
The general framework for averaging material properties is that described in \citet{Hill72}; further details can be found
in \citet{Nasser}.

Consider an aggregate of single crystals, each constituent being characterized by its orientation with respect
to the global Cartesian frame in the reference configuration of the aggregate and by its constitutive response as described above.
The response of the aggregate is to represent the constitutive response of a macro-particle
$\overline{X}$, which particle is subjected to the macro deformation gradient $\overline{\bb{F}}$.
With $\bb{F}(X,t)$ denoting the field of deformation gradient within the aggregate,
let $\bb{T} := \bb{\tau}\bb{F}^{-T} = \bb{F}\bb{S}$ denote the corresponding nonsymmetric Piola-Kirchhoff, or nominal stress field.
The motion of the aggregate is viewed as a sequence of equilibrium states which are governed by the following equations:
\begin{equation}
\makebox{Div\,}\bb{T}(X,t) = \bb{0},\,\,\, X\in\Omega
\label{Hbvp1}
\end{equation}
\begin{equation}
\bb{T}(X,t)=\bb{F}(X,t)\frac{\partial\phi}{\partial\bb{E}}\left(X,\bb{E},\bb{E}^p\right),\,\,\,X\in\Omega
\label{Hbvp2}
\end{equation}
\begin{equation}
\hat{x}(X,t) = \overline{\bb{F}}(t):X,\,\,\, X\in\partial\Omega
\label{Hbvp3}
\end{equation}
where $X$ is now an explicit argument of the elastic potential, since the constituents may be elastically anisotropic,
and $\Omega$ is the domain occupied by the aggregate in the reference configuration.
Under these conditions, the overall deformation gradient is the direct average of the corresponding field inside the RVE,
and the overall nominal stress $\overline{\bb{T}}$ is then defined as the direct average of the stress field:
\begin{equation}
\overline{\bb{F}} = \frac{1}{|\Omega|}\int_{\Omega}\bb{F}(X,t)\,dX,\,\,\,\,
\overline{\bb{T}}(t):=\frac{1}{|\Omega|}\int_{\Omega}\bb{T}(X,t)\,dX
\end{equation}
The local nominal stress field satisfies also the equilibrium of momentum of momentum in the form $\bb{F}\bb{T}^T=\bb{T}\bb{F}^T$.
Averaging this identity over the domain of the aggregate, and taking into account that $\bb{T}$ is equilibrated and
$\bb{F}$ is compatible (being the gradient of $\hat{x}$), by Hill's Lemma it follows $\overline{\bb{F}}\,\overline{\bb{T}}^T=
\overline{\bb{T}}\,\overline{\bb{F}}^T$.
Hill's Lemma applies also to the equilibrated and compatible fields $\bb{T}$ and $\bb{\dot{\bb{F}}}$, respectively,
and hence there holds
\begin{equation}
\frac{1}{|\Omega|}\int_{\Omega}\bb{T}\cdot\bb{\dot{F}}\,dX = \overline{\bb{T}}\cdot\dot{\overline{\bb{F}}}
\label{AverageTF}
\end{equation}
To $\overline{\bb{F}}$ one can associate an overall Green strain $\overline{\bb{E}}$ and a corresponding overall symmetric
(by the overall momentum of momentum equilibrium) Piola-Kirchhoff stress $\overline{\bb{S}}$  which is defined by requiring that the pairs  $\left(\overline{\bb{S}},\overline{\bb{E}}\right)$ and $\left(\overline{\bb{T}},\overline{\bb{F}}\right)$ be work-conjugated:
\begin{equation}
\overline{\bb{E}}:=\frac{1}{2}\left(\overline{\bb{F}}^T\overline{\bb{F}}-\bb{I}\right),\,\,\,\,
\overline{\bb{S}}:=\overline{\bb{F}}^{(-1)}\overline{\bb{T}}
\end{equation}

\subsection{Macro elastic potential and macro rate of plastic deformation}

\noindent
By definition, the aggregate is subject to elastic deformation if all of its constituents experience elastic deformation.
Assume the current stress state is elastic; then given an arbitrary overall direction of motion $\dot{\overline{\bb{F}}}$,
there exists a time interval, no matter how small, during which the deformation in this direction remains elastic.
Then, with $\bb{S}$ representing the symmetric P-K stress field within the aggregate and $\phi$ the elastic potential,
taking into account that the plastic state of the aggregate does not vary during an elastic process, we have, by employing eq.(\ref{AverageTF}):
\begin{equation}
\hspace{-1cm}
\overline{\bb{S}}\cdot\dot{\overline{\bb{E}}} = \overline{\bb{T}}\cdot\dot{\overline{\bb{F}}} =
\frac{1}{|\Omega|}\int_{\Omega}\bb{T}\cdot\bb{\dot{F}}\,dX = \frac{1}{|\Omega|}\int_{\Omega}\bb{S}\cdot\bb{\dot{E}}\,dX
=\frac{d}{dt}\frac{1}{|\Omega|}\int_{\Omega}\phi\left(X,\bb{E},\bb{E}^p\right)\,dX
\label{MacroPotential0}
\end{equation}
Following \citet{HillRice73} we define
\begin{equation}
\overline{\phi}\left(\overline{X},\overline{\bb{E}},\bb{H}\right):=\frac{1}{|\Omega|}\int_{\Omega}\phi\left(X,\bb{E},\bb{E}^p\right)\,dX
= \frac{1}{|\Omega|}\int_{\Omega}\phi^e\left(X,\bb{E}-\bb{E}^p\right)\,dX
\label{MacroPotential}
\end{equation}
The third argument of $\overline{\phi}$, $\bb{H}$,  may consist of any set of variables that
parameterize the plastic state at the continuum particle $\overline{X}$.
It is assumed that sufficient conditions are met by the local elastic potential $\phi$ so that the local deformation gradient $\bb{F}$
is uniquely determined, via BVP(\ref{Hbvp1})-(\ref{Hbvp3}), by the overall deformation gradient $\overline{\bb{F}}$, see, for example, \citet{MarsdenHughes} for a discussion on existence and uniqueness of solutions in the context of finite elasticity.
Then with $\overline{\bb{F}} = \overline{\bb{R}}\,\overline{\bb{U}}$ representing the polar decomposition of the macro deformation gradient,
and $\bb{B}(X,\overline{\bb{U}},\bb{H})$ the deformation gradient of the solution of BVP(\ref{Hbvp1})-(\ref{Hbvp3}) with $\hat{x}(X,t)=\overline{\bb{U}}:X$ on the boundary $\partial\Omega$, the local deformation gradient $\bb{F}$ corresponding to uniform $\overline{\bb{F}}$ on the boundary admits the representation
\begin{equation}
\bb{F}(X,t) = \overline{\bb{R}}\,\bb{B}(X,\overline{\bb{U}},\bb{H})\,\,\Longrightarrow\,\, \bb{E}(X,t) = \bb{M}(X,\overline{\bb{E}},\bb{H})
\label{Moperator}
\end{equation}
where, above, it was defined $\bb{M} := (\bb{B}^T\bb{B}-\bb{I})/2$ and the relationship
$\overline{\bb{U}} = (\bb{I}+2\overline{\bb{E}})^{1/2}$ was considered.
Hence the local strain field $\bb{E}$ is uniquely determined by the macro strain $\overline{\bb{E}}$, justifying the adoption of the second argument of $\overline{\phi}$.
Combining eqs.(\ref{MacroPotential0}) and (\ref{MacroPotential}), recalling that the current deformation is elastic
and that $\dot{\overline{\bb{E}}}$ is arbitrary, it follows that
\begin{equation}
\overline{\bb{S}}\cdot\dot{\overline{\bb{E}}} = \frac{\partial\overline{\phi}}{\partial\overline{\bb{E}}}\left(\overline{\bb{E}},\bb{H}\right)\cdot\dot{\overline{\bb{E}}}
\,\,\,\Longrightarrow\,\,\, \overline{\bb{S}} = \frac{\partial\overline{\phi}}{\partial\overline{\bb{E}}}\left(\overline{\bb{E}},\bb{H}\right)
\label{MacroPotentialA}
\end{equation}
showing that the elastic response of the aggregate is hyper-elastic with the macro potential defined in eq.(\ref{MacroPotential}).
For simplicity, the macro-particle $\overline{X}$ will be dropped in what follows.

The overall rate of plastic deformation can now be defined by
employing the rate version of the stress-strain relationship.
Thus, taking the time rate of the stress-stress relationship in eq.(\ref{MacroPotentialA})
along a general path of deformation, involving plastic flow within the aggregate,
there follows:
\begin{equation}
\dot{\overline{\bb{S}}} = \overline{\bb{K}}:\left[\overline{\bb{D}}-\overline{\bb{K}}^{(-1)}\overline{\bb{K}}^p:\bb{\dot{H}}\right] =
\overline{\bb{K}}:\left[\overline{\bb{D}}-\overline{\bb{D}}^{pL}\right]
\label{MCrystalRateForm}
\end{equation}
where $\overline{\bb{D}} := \dot{\overline{\bb{E}}}$ and:
\begin{equation}
\overline{\bb{K}}:=\frac{\partial^2\overline{\phi}}{\partial\overline{\bb{E}}\partial\overline{\bb{E}}},\,\,\,\,
\overline{\bb{K}}^p:=-\frac{\partial^2\Phi}{\partial\overline{\bb{E}}\partial\bb{H}},\,\,\,\,
\overline{\bb{D}}^{pL} := \overline{\bb{K}}^{(-1)}\overline{\bb{K}}^p:\bb{\dot{H}}
\label{MCrystalRateForm2}
\end{equation}
We call $\overline{\bb{D}}^{pL}$ the overall rate of plastic deformation.

Regarding the parametrization of the plastic state of the aggregate, we have, on one hand,
\begin{displaymath}
\frac{d}{dt}\overline{\phi}\left(\overline{\bb{E}},\bb{H}\right) =
\frac{1}{|\Omega|}\int_{\Omega}\frac{\partial\phi^e}{\partial\bb{E}^e}\cdot\left(\bb{\dot{E}}-\bb{\dot{E}}^p\right)dX=
\frac{1}{|\Omega|}\int_{\Omega}\bb{S}\cdot\left(\bb{D}-\widehat{\bb{D}}^{pL}\right)dX
\end{displaymath}
and on the other
\begin{displaymath}
\frac{d}{dt}\overline{\phi}\left(\overline{\bb{E}},\bb{H}\right) =
\frac{\partial\overline{\phi}}{\partial\overline{\bb{E}}}\cdot\dot{\overline{\bb{E}}} +
\frac{\partial\overline{\phi}}{\partial\bb{H}}\cdot\bb{\dot{H}} = \overline{\bb{S}}\cdot\overline{\bb{D}}+\frac{\partial\overline{\phi}}{\partial\bb{H}}\cdot\bb{\dot{H}}
\end{displaymath}
from where it follows, by recalling that $\overline{\bb{S}}\cdot\overline{\bb{D}} = (1/|\Omega|)\int_{\Omega}\bb{S}\cdot\bb{D}\,dX$,
\begin{equation}
\frac{\partial\overline{\phi}}{\partial\bb{H}}\cdot\bb{\dot{H}} = \frac{-1}{|\Omega|}\int_{\Omega}\bb{S}\cdot\widehat{\bb{D}}^{pL}\,dX
\label{PWork1}
\end{equation}
Eqs.(\ref{MacroPotentialA}), (\ref{MCrystalRateForm2}) and (\ref{PWork1}) are the basis for the specification
of the parameters collected under the abstract notation $\bb{H}$.
An illustration will be instructive, since it will allow us to make contact with the traditional
approach to parameterizing the plastic macro-state.
However, to make the calculations explicit the general homogenization context will be momentarily relaxed.

\noindent
\textbf{Illustration.} \citet{Taylor38}'s homogenization
principle states, based on compatibility reasons, that the
deformation is uniform within each constituent and equal to the
deformation at the boundary of the aggregate. In the present
formalism, this amounts to assuming that
$\bb{M}(X,\overline{\bb{E}},\bb{H}) = \overline{\bb{E}}$, $\forall
X\in\Omega$, thus disregarding any equilibrium considerations
within the aggregate. Then, by eqs.(\ref{MacroPotentialA}) and
(\ref{Moperator}),
\begin{equation}
\begin{array}{lll}
\dd\overline{\bb{S}}
&=&\dd\frac{1}{|\Omega|}\int_{\Omega}\left(\frac{\partial\bb{M}}{\partial\overline{\bb{E}}}\right)^T:\frac{\partial\phi}{\partial\bb{E}}(X,\bb{E},\bb{E}^p)dX
=\vp\\
\dd&=&\dd\frac{1}{|\Omega|}\int_{\Omega}\frac{\partial\phi}{\partial\bb{E}}(X,\bb{E},\bb{E}^p)dX = \frac{1}{|\Omega|}\int_{\Omega}\bb{S}\,dX
\end{array}
\label{TaylorStress}
\end{equation}
Also, for simplicity, assume that the local elastic potential is the quadratic\\
$\phi^{e}(\bb{E}-\bb{E}^p)=(1/2)\left[\bb{K}:\left(\bb{E}-\bb{E}^p\right)\right]\cdot\left(\bb{E}-\bb{E}^p\right)$.
Then straightforward calculations, starting from eq.(\ref{MacroPotential}), obtain
\begin{equation}
\dd\overline{\phi}\left(\overline{\bb{E}},\bb{H}\right) = \frac{1}{2}\left(\overline{\bb{K}}:\overline{\bb{E}}\right)\cdot\overline{\bb{E}}
-\left(\overline{\bb{K}}:\overline{\bb{E}}^p\right)\cdot\overline{\bb{E}}+\frac{1}{2}\overline{e}^p
\label{Macroquad}
\end{equation}
where
\begin{displaymath}
\begin{array}{l}
\dd\overline{\bb{K}}:=\frac{1}{|\Omega|}\int_{\Omega}\bb{K}\,dX,\,\,\,\,\,
\overline{\bb{E}}^{p}:=\frac{1}{|\Omega|}\int_{\Omega}\overline{\bb{K}}^{(-1)}\bb{K}:\bb{E}^{p}\,dX,\vp\\
\dd \makebox{and}\,\,\,\,\,
\overline{e}^p:= \frac{1}{|\Omega|}\int_{\Omega}\left(\bb{K}:\bb{E}^p\right)\cdot\bb{E}^p\,dX
\end{array}
\end{displaymath}
We show that $\bb{H}:=\left\{\overline{\bb{E}}^p, \overline{e}^p\right\}$ is a valid parametrization of the plastic state
of the aggregate. First, let us notice that $\overline{\bb{E}}^p$ and $\overline{e}^p$ are independent entities
(for example, $\overline{e}^p$ may be constant in situations where $\overline{\bb{E}}^p$ varies).
Then, by eqs.(\ref{TaylorStress}-\ref{Macroquad}):
\begin{displaymath}
\begin{array}{lll}
\dd\overline{\bb{S}} &=&\dd\frac{1}{|\Omega|}\int_{\Omega}\frac{\partial\phi^e}{\partial\bb{E}}\,dX = \frac{1}{|\Omega|}\int_{\Omega}\bb{K}:\left(\overline{\bb{E}}-\bb{E}^{p}\right)dX =
\overline{\bb{K}}:\left(\overline{\bb{E}}-\overline{\bb{E}}^p\right) =\vp\\
&=&\dd\frac{\partial\overline{\phi}}{\partial\overline{\bb{E}}}\left(\overline{\bb{E}},\overline{\bb{E}}^p,\overline{e}^p\right)
\end{array}
\end{displaymath}
and hence eq.(\ref{MacroPotentialA}) is verified. Next, by eq.(\ref{MCrystalRateForm2}) and eq.(\ref{MacroDP2}) below,
there must hold:
\begin{displaymath}
\overline{\bb{K}}^{(-1)}\overline{\bb{K}}^p:\bb{\dot{H}} = \overline{\bb{D}}^{pL} =
\frac{1}{|\Omega|}\int_{\Omega}\overline{\bb{K}}^{(-1)}\bb{K}:\bb{\dot{E}}^p\,dX \iff
\overline{\bb{K}}^p:\bb{\dot{H}} = \overline{\bb{K}}:\dot{\overline{\bb{E}}}^p
\end{displaymath}
Recalling the definitions in eq.(\ref{MCrystalRateForm2}), one has $\overline{\bb{K}}^p = \left[\overline{\bb{K}},\bb{0}\right]$,
and $\bb{\dot{H}} = \left\{\dot{\overline{\bb{E}}}^p, \dot{\overline{e}}^p\right\}$; hence the above identity is also satisfied.
Finally,
\begin{displaymath}
\begin{array}{lll}
\dd\frac{\partial\overline{\phi}}{\partial\overline{\bb{E}}^p}\cdot\dot{\overline{\bb{E}}}^p+
\frac{\partial\overline{\phi}}{\partial\overline{e}^p}\,\dot{\overline{e}}^p &=&\dd
-\left(\overline{\bb{K}}:\overline{\bb{E}}\right)\cdot\dot{\overline{\bb{E}}}^p+
\frac{1}{|\Omega|}\int_{\Omega}\left(\bb{K}:\bb{E}^p\right)\cdot\bb{\dot{E}}^{p}\,dX =\vp\\
&=&\dd\frac{-1}{|\Omega|}\int_{\Omega}\bb{S}\cdot\widehat{\bb{D}}^{pL}\,dX
\end{array}
\end{displaymath}
and hence the identity in eq.(\ref{PWork1}) is verified, thus
closing our proof. $\square$

Some of the features of the above example extend to the general
case. Indeed, having defined a macro-rate of plastic deformation,
by eqs.(\ref{MCrystalRateForm}-\ref{MCrystalRateForm2}), one can
always associate with it a measure of plastic deformation in the
form of the symmetric second order tensor
$\overline{\bb{E}}^p(X,t)$ solution of the Cauchy problem
\begin{equation}
\dot{\overline{\bb{E}}}^p =
\overline{\bb{D}}^{pL}\,\,\,\,\makebox{with}\,\,\,\,\,\overline{\bb{E}}^p(X,t_0)
= \bb{0}
\end{equation}
Then a parametrization of the plastic state may be given in the
form $\bb{H} = \left\{\overline{\bb{E}}^p,\bb{Q}\right\}$, where
$\bb{Q}$ denotes a set of any additional plastic parameters. With
$\overline{\phi}=\overline{\phi}\left(\bb{E},\overline{\bb{E}}^p,\bb{Q}\right)$,
eq.(\ref{PWork1}) rewrites:
\begin{displaymath}
\frac{\partial\overline{\phi}}{\partial\overline{\bb{E}}^p}\cdot\dot{\overline{\bb{E}}}^p
+ \frac{\partial\overline{\phi}}{\partial\bb{Q}}\cdot\bb{\dot{Q}}=
\frac{-1}{|\Omega|}\int_{\Omega}\bb{S}\cdot\widehat{\bb{D}}^{pL}\,dX
\end{displaymath}
Formally, nothing seems to prevent the consideration of a macro
stress potential of the form
$\overline{\phi}\left(\bb{E},\overline{\bb{E}}^p,\bb{Q}\right) =
\widehat{\phi}\left(\overline{\bb{E}}^e,\bb{Q}\right)$, with
$\overline{\bb{E}}^e:=\overline{\bb{E}}-\overline{\bb{E}}^p$ and
possibly different characterizations of the $Q$-parameters of the
$\widehat{\phi}$ function. However, in this case the presence of
the additional parameters $\bb{Q}$ is mandatory. Indeed, if these
were absent, then due to the relationship
$\partial\overline{\phi}/\partial\overline{\bb{E}}^p = -
\partial\widehat{\phi}/\partial\overline{\bb{E}}^e =
-\overline{\bb{S}}$, the above identity would reduce to
\begin{displaymath}
\overline{\bb{S}}\cdot\overline{\bb{D}}^{pL} =
\frac{1}{|\Omega|}\int_{\Omega}\bb{S}\cdot\widehat{\bb{D}}^{pL}\,dX
\end{displaymath}
This dissipation identity is false, in general; by
eq.(\ref{MacroDP2}) below, an identity of this type holds, in
general, only for elastic directions.


\subsection{Macro yield surface}
\label{MacroYSurface}

\noindent
The elastic domain at the continuum particle $\overline{X}$ is defined as the set of all macro-stresses $\overline{\bb{S}}^e$ that can be reached
from the current stress $\overline{\bb{S}}$ by elastic deformation of the aggregate, \cite{Hill67}.
The overall stress associated with an elastic process starting at $\overline{\bb{F}}$, the deformation gradient at the current moment $t$, and ending at $\overline{\bb{F}}^e$ is
\begin{equation}
\overline{\bb{S}}^e = \frac{\partial\overline{\phi}}{\partial\overline{\bb{E}}}\left(\overline{\bb{E}}^e,\bb{H}\right)
\end{equation}
with $\overline{\bb{E}}^e$ denoting the Green strain associated with $\overline{\bb{F}}^e$.
It is assumed that the single crystal potential is strictly convex and co-finite, \citet{Rockafellar},
so that by eq.(\ref{MacroPotential}) the overall potential $\overline{\phi}$ is also strictly convex and co-finite.
Under these conditions the above stress-strain relationship can be inverted in the form
\begin{equation}
\overline{\bb{E}}^e = \frac{\partial\overline{\psi}}{\partial\overline{\bb{S}}}
\left(\overline{\bb{S}}^e,\bb{H}\right),\makebox{\,\,\,\,with\,\,\,\,}
\overline{\psi}\left(\overline{\bb{S}},\bb{H}\right):=\overline{\bb{S}}\cdot\overline{\bb{E}}-
\overline{\phi}\left(\overline{\bb{E}},\bb{H}\right)
\label{OVE}
\end{equation}
The local stress field $\bb{S}^e$ can now be calculated in terms of the overall stress $\overline{\bb{S}}^e$
via eqs.(\ref{CrystalRateForm}) and (\ref{Moperator}).
Then eq.(\ref{GeneralSchmid2}) leads to the following set of inequalities to be satisfied by $\overline{\bb{S}}^e$
\begin{equation}
\hspace{-1cm}
\begin{array}{l}
\dd\frac{\partial\phi}{\partial\bb{E}}\left(\bb{M}\left(X,\overline{\bb{E}}^e,\bb{H}\right),\bb{H}\right)
\cdot\bb{A}^{\alpha}
+
\hat{\zeta}^{\alpha}
\left(\frac{\partial\phi}{\partial\bb{E}}\left(\bb{M}\left(X,\overline{\bb{E}}^e,\bb{H}\right),\bb{H}\right),\bb{P}\right)<\vspace{2pt}\\
\dd<\tau_{cr}^{\alpha}\left[1+\hat{\zeta}^{\alpha}_h
\left(\frac{\partial\phi}{\partial\bb{E}}\left(\bb{M}\left(X,\overline{\bb{E}}^e,\bb{H}\right),\bb{H}\right),\bb{P}_h\right)\right]
\end{array}
\label{ElastDomain}
\end{equation}
for all $\alpha\in\mathcal{S}$ and every $X\in\Omega$, where $\overline{\bb{E}}^e$ is
given by eq.(\ref{OVE}) and $\bb{A}^{\alpha}$ also depend on the current stress state via eq.(\ref{HatDP}).
In general, it cannot be expected that the set defined by each of the above
inequalities be convex, see Appendix for a counter-example.
On the other hand, deviations from convexity are so small that one may assume, without much loss in rigor,
that the intersection of the sets defined by the above set of inequalities is convex.

Based on the above inequalities, it will be assumed that the macro elastic domain
can be characterized by employing a yield function $f$ in the form
\begin{equation}
f\left(\overline{\bb{S}}^e,...\right) = g_{\tau}\left(\overline{\bb{S}}^e,...\right)+g_l\left(\overline{\bb{S}}^e,...\right) - h(...)g_r\left(\overline{\bb{S}}^e,...\right)<0
\label{ElastDomain2}
\end{equation}
where $g\left(\overline{\bb{S}}^e,...\right):=g_{\tau}\left(\overline{\bb{S}}^e,...\right)+g_l\left(\overline{\bb{S}}^e,...\right)$ is to be representative of the left-hand side of the inequalities in eq.(\ref{ElastDomain}),
and referred to as the equivalent macro-stress; $g_{\tau}\left(\overline{\bb{S}}^e,...\right)$ is representative
of the left-hand side if the contributions of $\hat{\zeta}^{\alpha}$ in eq.(\ref{ElastDomain}) were absent;
$h(...)g_r\left(\overline{\bb{S}}^e,...\right)$ is to represent the right-hand side, being referred to as the hardening part of the macro yield function.
Only the macro-stress has been shown explicitly as argument since this will be relevant for our next developments;
the dots are to represent any additional (structural) parameters that may be required
for characterizing the shape and symmetries of the macro elastic domain, or its size.
The macro yield surface is by definition the boundary of the macro elastic domain; it is characterized
by $f\left(\overline{\bb{S}}^e,...\right) = 0$.

\subsection{Macro flow rule}

\noindent
Consider an arbitrary elastic direction of motion, originating at the current moment $t$;
this is a velocity field $\delta\hat{x}$ for which there exists a time interval such that continuing the motion
in this direction would induce an elastic state within the aggregate, \citet{Soare2013}.
It is characterized by the following BVP, obtained by linearizing BVP(\ref{Hbvp1})-(\ref{Hbvp3}) at $t$ with fixed plastic state:
\begin{equation}
\left\{
\begin{array}{l}
\dd\makebox{Div\,}\delta\bb{T}(X) = \bb{0},\,\,\, X\in\Omega\vp\\
\delta\bb{T}(X)=\left[\delta\bb{F}(X)\right]\bb{S}(X,t)+\bb{F}(X,t)\bb{K}:\delta\bb{E}(X),\,\,\,X\in\Omega\vp\\
\delta\hat{x}(X) = \delta\overline{\bb{F}}:X,\,\,\, X\in\partial\Omega
\end{array}
\right.
\label{HbvpL3}
\end{equation}
Corresponding macro elastic directions are defined as:
\begin{equation}
\delta\overline{\bb{E}}:=
\frac{1}{2}\left[\left(\delta\overline{\bb{F}}\right)^T\overline{\bb{F}}(t)+\overline{\bb{F}}^T(t)\delta\overline{\bb{F}}\right],\,\,\,
\delta\overline{\bb{S}}:=\overline{\bb{K}}:\delta\overline{\bb{E}}
\end{equation}
with $\overline{\bb{K}}$ defined in eq.(\ref{MCrystalRateForm2}).
During the virtual (or $\theta$-)motion relationship (\ref{Moperator}) continues to hold and hence,
by taking the $\theta$-time rate, at $\theta=0$, the local strain and stress fields are determined by the macro strain and stress increments as follows:
\begin{equation}
\delta\bb{E}(X) = \bb{P}:\delta\overline{\bb{E}}\,\,\,\Longrightarrow\,\,\,
\delta\bb{S} = \bb{K}:\delta\bb{E} = \bb{K}\bb{P}:\delta\overline{\bb{E}} = \bb{K}\bb{P}\overline{\bb{K}}^{(-1)}:\delta\overline{\bb{S}},
\label{Moperator2}
\end{equation}
where
\begin{equation}
\bb{P}(X,\overline{\bb{E}},\bb{H}):=\frac{\partial\bb{M}}{\partial\overline{\bb{E}}}(X,\overline{\bb{E}},\bb{H})
\end{equation}
Next, by employing the local stress-strain relationship in eq.(\ref{CrystalRateForm}) and the symmetry of $\bb{K}$:
\begin{equation}
\hspace{-1cm}
\delta\bb{S}\cdot\bb{D}^{pL} = \delta\bb{S}\cdot\left(\bb{D}-\bb{K}^{-1}:\bb{\dot{S}}\right)
=\delta\bb{S}\cdot\bb{D}-\left(\bb{K}^{-1}:\delta\bb{S}\right)\cdot\bb{\dot{S}}
=\delta\bb{S}\cdot\bb{D}-\delta\bb{E}\cdot\bb{\dot{S}}
\label{HillInv1}
\end{equation}
The last member in the above sequence of equalities is an instance of Hill's differential form,
\citet{Hill68b}, \citet{Hill72}. This form has the remarkable property that is invariant to changes
of conjugate stress-strain measures. For our purpose, the conjugate
pair $(\bb{T},\bb{F})$ is of primary interest; it is related to the pair $(\bb{S},\bb{E})$ by:
$\bb{T}=\bb{F}\bb{S}\,\Longrightarrow\, \delta\bb{T}=(\delta\bb{F})\bb{S}+\bb{F}\delta\bb{S},\,\,\makebox{and}\,\,
\bb{\dot{T}} = \bb{\dot{F}}\bb{S}+\bb{F}\bb{\dot{S}}$\,;
$\bb{E}=(1/2)\left(\bb{F}^T\bb{F}-\bb{I}\right)\,\Longrightarrow\,\delta\bb{E}=(1/2)\left[(\delta\bb{F}^T)\bb{F}+\bb{F}^T\delta\bb{F}\right],
\,\,\,\bb{D} = \bb{\dot{E}} = (1/2)\left(\bb{\dot{F}}^T\bb{F}+\bb{F}^T\bb{\dot{F}}\right)$.
With these relationships the following identity is easily verified:
\begin{equation}
\delta\bb{S}\cdot\bb{D}-\bb{\dot{S}}\cdot\delta\bb{E}=\delta\bb{T}\cdot\bb{\dot{F}}-\bb{\dot{T}}\cdot\delta\bb{F}
\label{HillInv2}
\end{equation}
Now $(\delta\bb{T},\bb{\dot{F}})$ and $(\bb{\dot{T}},\delta\bb{F})$ are pairs of equilibrated and compatible fields
and hence an average relationship like eq.(\ref{AverageTF}) holds for each.
Then averaging in (\ref{HillInv1}) and employing (\ref{HillInv2}) there follows
\begin{equation}
\hspace{-1cm}
\frac{1}{|\Omega|}\int_{\Omega}\delta\bb{S}\cdot\bb{D}^{pL}\,dX =
\left(\delta\overline{\bb{T}}\right)\cdot\dot{\overline{\bb{F}}} - \dot{\overline{\bb{T}}}\cdot\delta\overline{\bb{F}}
=\left(\delta\overline{\bb{S}}\right)\cdot\overline{\bb{D}} - \dot{\overline{\bb{S}}}\cdot\delta\overline{\bb{E}}
=\delta\overline{\bb{S}}\cdot\left(\overline{\bb{D}}-\overline{\bb{K}}^{(-1)}:\dot{\overline{\bb{S}}}\right)
\end{equation}
By eq.(\ref{MCrystalRateForm}), the bracketed term on the right of the last equality
above is the macro-rate of plastic deformation, $\overline{\bb{D}}^{pL}$.
Then, by substituting in the integral term, above, the relationship in eq.(\ref{Moperator2}) between the local and macro
elastic stress directions, one obtains an identity valid for arbitrary $\delta\overline{\bb{S}}$, leading
to the following representation of the macro-rate of plastic deformation:
\begin{equation}
\overline{\bb{D}}^{pL} = \frac{1}{|\Omega|}\int_{\Omega}\overline{\bb{K}}^{(-1)}\bb{P}^T\bb{K}:\bb{D}^{pL}\,dX
\label{MacroDP2}
\end{equation}

Averaging over the domain of the aggregate the inequality in eq.(\ref{NormalityRef}), and employing eqs.(\ref{MacroDP2}) and (\ref{Moperator2}), results in the inequality
\begin{equation}
\delta\overline{\bb{S}}\cdot\left(\overline{\bb{D}}^{pL}+\overline{\bb{Z}}\right)\leq 0,\,\,\,\,\,
\makebox{with}\,\,\,\,\,
\overline{\bb{Z}}:=
\frac{1}{|\Omega|}\int_{\Omega}\overline{\bb{K}}^{(-1)}\bb{P}^T\bb{K}:\bb{Z}\,dX
\label{MPFlowrule}
\end{equation}
valid for any macro elastic direction $\delta\overline{\bb{S}}$.
Assuming the macro yield function is smooth, 
the above inequality then translates into the following flow rule
\begin{equation}
\overline{\bb{D}}^{pL}+\overline{\bb{Z}} = \dot{\lambda}\frac{\partial f}{\partial\overline{\bb{S}}}\left(\overline{\bb{S}},...\right)
\label{ZFlowRule}
\end{equation}
with $\dot{\lambda}$ denoting a scalar parameter characterizing the magnitude of $\overline{\bb{D}}^{pL}+\overline{\bb{Z}}$.

\subsection{Eulerian description}

\noindent
Since many of the current investigations in metal plasticity employ a hypo-elastic
formulation of the stress-strain response, it is perhaps not
without relevance to rephrase the essence of the above results in
terms of spatial objects (defined on the current configuration of the continuum body).

With $\bb{S}$ and $\bb{\tau}:=J\bb{\sigma}$ (re)denoting the symmetric P-K and, respectively, the Kirchhoff macro-stresses, the stress-strain
relationship, in the rate form of eq.(\ref{MCrystalRateForm}), reads, in the reference and current configurations:
\begin{equation}
\hspace{-1cm}
\bb{\dot{S}} = \bb{K}:\left[\bb{D}-\bb{D}^{pL}\right]\,\,\iff\,\, \bb{\dot{\tau}}^L = \bb{K}^L:\left[\bb{d}-\bb{d}^{pL}\right]
\,\,\iff\,\, \bb{\dot{\tau}}^J = \bb{K}^J:\left[\bb{d}-\bb{d}^{p}\right]
\label{SpatialPrates}
\end{equation}
where $\bb{d}$ is the eulerian macro-rate of deformation, i.e. the symmetric part of the spatial velocity gradient $\bb{l}$,
$\bb{F}$ the macro deformation gradient,
$\bb{D}=\bb{F}^T\bb{d}\bb{F}$,  $\bb{D}^{pL}=\bb{F}^T\bb{d}^{pL}\bb{F}$,
$\bb{K}^L:\bb{a} = \bb{F}[\bb{K}:(\bb{F}^T\bb{a}\bb{F})]\bb{F}^T$, for any symmetric $\bb{a}$, $\bb{\dot{\tau}}^L = \bb{\dot{\tau}}-\bb{l}\bb{\tau}-\bb{\tau}\bb{l}^T$, $\bb{\dot{\tau}}^J = \bb{\dot{\tau}}+\bb{\tau}\bb{w}-\bb{w}\bb{\tau}$,
$\bb{w}:=(\bb{l}-\bb{l}^T)/2$, $\bb{K}^J:\bb{a} = \bb{K}^L:\bb{a}+\bb{\tau}\bb{a}+\bb{a}\bb{\tau}$, for any symmetric $\bb{a}$,
and $\bb{d}^p:=\left(\bb{K}^J\right)^{-1}\bb{K}^L:\bb{d}^{pL}$.

In the context of Section \ref{MacroYSurface}, let $\Delta\bb{F}^e = \bb{F}^e\bb{F}^{-1}$ denote
the macro deformation gradient of an arbitrary macro elastic deformation starting from the current configuration defined by
the macro deformation gradient $\bb{F}$.
With the polar decomposition $\Delta\bb{F}^e = \bb{R}\bb{U}$, for any elastic stress state $\bb{S}^e$ there holds
$\bb{S}^e = \bb{F}^{e-1}\bb{\tau}\bb{F}^{e-T} = \bb{F}^{-1}\bb{U}^{-1}\bb{\tau}^R\bb{U}^{-1}\bb{F}^{-T}$, where
$\bb{\tau}^R:=\bb{R}^T\bb{\tau}\bb{R}$. Since the deformation is elastic,
$\bb{S}^e$ and $\bb{U}$ are in a one-to-one relationship and hence $\bb{\tau}^R$ and $\bb{S}^e$ also are in a one-to-one relationship.
One can define then $f^R(\bb{\tau}^R,...):=f^S\left(\bb{F}^{-1}\bb{U}^{-1}\bb{\tau}^R\bb{U}^{-1}\bb{F}^{-T},...\right)$,
$f^S$ now denoting the overall yield surface defined in eq.(\ref{ElastDomain2}).
Letting $\bb{Q}$ denote the rotation associated with the material spin, $\bb{\dot{Q}}=\bb{w}\bb{Q}$, a further definition
of the macro yield function written with respect to axes that spin with $\bb{w}$,
$f^Q(\bb{\tau}^Q,...):=f^R(\bb{\tau}^R,...)$, allows us to characterize the elastic domain in the form $f^Q(\bb{\tau}^Q,..)<0$;
also, by defining $f(\bb{\tau},...):=f^Q(\bb{\tau}^Q,...)$, one can reformulate this characterization in terms of the Kirchhoff stress itself by writing $f(\bb{\tau},...)<0$. In the latter case it is implied that, for anisotropic yielding properties,
the arguments of $f$ include some characteristic structural tensors, e.g., \citet{Liu}, \citet{Boehler},
for otherwise the principle of objectivity would restrict $f$ to an isotropic function.

With $f^Q$ describing the elastic domain, its elastic directions, issued at the current stress $\bb{\tau}$, are of the form $\delta\bb{\tau}^Q = \bb{Q}^T(\delta\bb{\tau}+\bb{\tau}\bb{w}-\bb{w}\bb{\tau})\bb{Q} = \bb{Q}^T(\delta^J\bb{\tau})\bb{Q}$, the latter equality serving as definition of the operator $\delta^J$. Along elastic directions there holds $\delta^L\bb{\tau} = \bb{K}^L:\bb{d}$ and then $\delta^J$ is related to $\delta^L\bb{\tau}:=\delta\bb{\tau}-\bb{l}\bb{\tau}-\bb{\tau}\bb{l}^T$
by $\delta^J\bb{\tau} = \bb{N}:\delta^L\bb{\tau}$, with $\bb{N}:\bb{a}=\bb{a}+\bb{\tau}(\bb{K}^L)^{-1}:\bb{a}+[(\bb{K}^L)^{-1}:\bb{a}]\bb{\tau}$,
for any symmetric second order tensor $\bb{a}$.
The $\bb{N}$-operator relates also the two spatial plastic rates in eq.(\ref{SpatialPrates}) by $\bb{d}^{pL} = \bb{N}^T:\bb{d}^p$.
Then, with eq.(\ref{MPFlowrule}) there holds
\begin{displaymath}
\hspace{-1cm}
0\geq\delta\bb{S}\cdot\left(\bb{D}^{pL}+\overline{\bb{Z}}\right) = \delta\bb{\tau}^L\cdot\left(\bb{d}^{pL}+\bb{z}^L\right)
=\delta^J\bb{\tau}\cdot\left(\bb{d}^{p}+\bb{z}\right) = \delta\bb{\tau}^Q\cdot\left[\bb{Q}^T\left(\bb{d}^{p}+\bb{z}\right)\bb{Q}\right]
\end{displaymath}
where $\bb{z}^L:=\bb{F}^{-T}\overline{\bb{Z}}\bb{F}^{-1}$ and $\bb{z}:=\bb{N}^{-T}:\bb{z}^L$.
With $\delta\bb{\tau}^Q$ arbitrary, from the last inequality above there follows
\begin{equation}
\bb{d}^{p}+\bb{z} = \dot{\lambda}\,\bb{Q}\,\frac{\partial f^Q}{\partial\bb{\tau}^Q}\,\bb{Q}^T = \dot{\lambda}\,\frac{\partial f}{\partial\bb{\tau}}
\label{ZFlowRule2}
\end{equation}
This is the Eulerian form of the flow rule in eq.(\ref{ZFlowRule}). When non-Schmid effects are absent,
it reduces to the classical normality rule.

Finally, let us note that by the definition of $\overline{\bb{Z}}$ and $\bb{Z}$ in eqs.(\ref{MPFlowrule}) and (\ref{NormalityRef}),
the deviation from the normal direction can be decomposed as $\bb{z} = \bb{z}_l+\bb{z}_r$,
where $\bb{z}$ is representative of $\zeta^{\alpha}$ (or "left-hand") effects,
whereas $\bb{z}_r$ is representative of $\zeta^{\alpha}_h$ (or "right-hand") effects;
also, by eq.(\ref{ElastDomain2}), the yield function admits the Eulerian representation
$f(\bb{\tau},...) = g_{\tau}(\bb{\tau},...)+g_l(\bb{\tau},...)-h(...)g_r(\bb{\tau},...)$.
Then the flow rule in eq.(\ref{ZFlowRule2}) can be further specified in the form:
\begin{equation}
\bb{d}^{p}+\bb{z}_l+\bb{z}_r = \dot{\lambda}\,
\left(\frac{\partial g_{\tau}}{\partial\bb{\tau}}+\frac{\partial g_l}{\partial\bb{\tau}}-
h\,\frac{\partial g_r}{\partial\bb{\tau}}\right)
\label{MPFlowruleSpace}
\end{equation}

\section{Conclusions}

\noindent
The overall response of a polycrystal, representative of a
continuum particle $X$, has been analyzed here under the following
conditions: 1) the stress-strain response of a constituent crystal
derives from an elastic potential; 2) the slip systems of a
constituent convect with the crystal lattice; 3) an extended
Schmid law, incorporating non-Schmid effects, characterizes slip
activity. Then a macro elastic potential
$\Phi=\Phi(\bb{E},\bb{H})$ exists, \citet{HillRice73}, such that the response at
particle $X$ is given by
\begin{equation}
\bb{S} = \frac{\partial\Phi}{\partial\bb{E}}(\bb{E},\bb{H})
\label{zzzPotential}
\end{equation}
where $\bb{S}$ and $\bb{E}$ are the symmetric Piola-Kirchhoff stress and the Green strain with respect to a reference
configuration. $\bb{H}$ is a collection of variables that parameterize the current plastic state at particle $X$.
A macro yield function exists and describes the elastic domain at particle $X$ via the inequality
$f(\bb{S},...)=g(\bb{S},...)-h(...)g_h(\bb{S},...)<0$, where $h(...)$ is a macro-measure of hardening.
The evolution of the plastic state is characterized by the associated flow rule
\begin{equation}
\bb{K}^{-1}\bb{K}^p:\bb{\dot{H}} +\bb{Z} = \dot{\lambda}\frac{\partial f}{\partial\bb{S}}\,,\,\,\,\,\,
\makebox{with}\,\,\,\,\bb{K}:=\frac{\partial^2\Phi}{\partial\bb{E}\partial\bb{E}}\,,\,\,\,
\bb{K}^p:=\frac{\partial^2\Phi}{\partial\bb{E}\partial\bb{H}}
\label{zzzAssociate}
\end{equation}
where $\dot{\lambda}$ is further determined, as usual, from the consistency condition.
$\bb{K}^{-1}\bb{K}^p:\bb{\dot{H}}$ is the macro-rate of plastic deformation and
$\bb{Z}$ is a symmetric second order tensor
representative of the non-Schmid effects manifesting at constituent level.

It may therefore be concluded that if the resolved shear stress on
a slip system is the main factor influencing the activity of that
slip system, then the overall direction of plastic deformation and
the exterior normal to the macro yield surface are always related
by an associated flow rule. When non-Schmid effects are present,
an additional macro-variable, $\bb{Z}$, enters additively the flow
rule, measuring the deviation of the overall rate of plastic
deformation from the stress-gradient of the macro yield surface.

\appendix
\section{An alternative parametrization of the plastic state}

\noindent
Starting from the multiplicative elastic-plastic decomposition in eq.(\ref{MultDec}) it is possible
to develop a different parametrization of the plastic state of a single crystal,
based on the supposedly more intuitive and hence a priori concept of elastic (or plastic) strain.
Its details and consequences at macro-level are examined next.

\subsection{An additive elastic-plastic decomposition}

\noindent
The context being that of eq.(\ref{MultDec}), the Green strain
\begin{equation}
\bb{E}^*:= \frac{1}{2}\left(\bb{C}^*-\bb{I}\right), \,\makebox{ with }\, \bb{C}^*:=\bb{F}^{*T}\bb{F}^*,
\label{Estar}
\end{equation}
is adopted as measure of the elastic strain between the intermediary and current configurations.
To define its image (or pull-back) $\bb{E}^e$ in the reference configuration, let $dX\in T_X$ be an arbitrary vector
with its image $dx=\bb{F}:dX = \bb{F}^*:dX^p$; with $dl_p$ and $dl$ denoting the lengths of $dX$ in the intermediate and current configurations,
\begin{equation}
\left(\bb{E}^e:dX\right)\cdot dX := (dl^2-dl_p^2)/2 = \left(\bb{E}^*:dX^p\right)\cdot dX^p
\label{Eelastic}
\end{equation}
With $dX^p=\bb{F}^p:dX$, there follows:
\begin{equation}
\hspace{-1cm}
\bb{E}^e=\bb{F}^{pT}\bb{E}^*\bb{F}^p = \frac{1}{2}(\bb{C}-\bb{C}^p) = \bb{E}-\bb{E}^p,\,\,\makebox{ with }\, \bb{C}:=\bb{F}^T\bb{F},\,\, \bb{C}^p:=\bb{F}^{pT}\bb{F}^p,
\label{Edecomp}
\end{equation}
and the total and plastic Green strain tensors $\bb{E}$ and $\bb{E}^p$ defined by
\begin{equation}
\bb{E}:=\left(\bb{C}-\bb{I}\right)/2,\,\,\,
\bb{E}^p:=\left(\bb{C}^p-\bb{I}\right)/2
\end{equation}
The additive decomposition in eq.(\ref{Edecomp}), i.e. $\bb{E}=\bb{E}^e+\bb{E}^p$,
was postulated in the pioneering work of \citet{Green} on a  macro-plasticity theory at finite strain. By the above arguments,
it is equivalent with the multiplicative decomposition in eq.(\ref{MultDec}).
However, one may note that this decomposition differs from the one associated with the plastic parameter defined in eq.(\ref{PlasticStrain}).
One may also note that both
$\bb{E}^p$ and $\bb{E}^e$ are invariant to any orthogonal transformation of the intermediary configuration,
hence the advantage of using a Lagrangian formulation based on the reference state over a Lagrangian formulation based on
the intermediary configuration, the latter being prominent in many of the recent works on crystal plasticity; see also the discussion in \citet{GreenNaghdi71}.

$\bb{E}^p$ can now be employed to parameterize the plastic state of the crystal.
For example, an elastic potential often employed to model the elasticity of single crystals
is the quadratic in $\bb{E}^e$, leading to a "linear" stress-strain relationship
\begin{equation}
\phi(\bb{E}^e) = \frac{1}{2}\left(\bb{K}:\bb{E}^e\right)\cdot\bb{E}^e\Longrightarrow
\bb{S} = \bb{K}:\left(\bb{E}-\bb{E}^p\right)
\label{QuadraticP}
\end{equation}
with $\bb{K}$ being a constant symmetric and positive definite fourth order tensor, e.g., \citet{Kalidindi}, \citet{Anand96}, or \citet{Miehe}.
Certainly, eq.(\ref{QuadraticP}) reproduces the basic features of small strain elasticity, appropriate for metals,
when $\bb{E}^e$ is in a small vicinity of $\bb{0}$.
It is remarked that in the cited works, $\bb{E}^*$, as defined by eq.(\ref{Estar}), is used as primary elastic strain measure, and the symmetric Kirchhoff stress $\bb{S}^*:=\bb{F}^{*-1}J^*\bb{\sigma}\bb{F}^{*-T}$ as stress measure in the intermediate configuration,
to define the elastic law in the form $\bb{S}^* = \bb{K}:\bb{E}^*$.
This relationship can be obtained by the push-forward of eq.(\ref{QuadraticP}) to the
intermediate configuration, while assuming, formally, that $\bb{K}$ is left unchanged by plastic flow, that is, by $\bb{F}^p$

Then the rate of plastic deformation takes the form
\begin{equation}
\bb{D}^{pL}:=\bb{\dot{E}}^p = \bb{F}^p\frac{1}{2}\left(\bb{L}^p+\bb{L}^{pT}\right)\bb{F}^p
\label{DPref}
\end{equation}
Of further interest is its relationship with the rate $\widehat{\bb{D}}^{pL}$ defined in eq.(\ref{DPrefTruesdell2}).
To deduce it, one may first notice that
\begin{equation}
\hspace{-0.5cm}
\bb{D}^e:=\bb{\dot{E}}^e = \bb{D}^*+\bb{D}^z, \,\,\,\,\makebox{where}\,\,\,\,
\bb{D}^z:=\bb{F}^{pT}\bb{L}^{pT}\bb{F}^{p-T}\bb{E}^e+\bb{E}^e\bb{F}^{p-1}\bb{L}^p\bb{F}^p
\end{equation}
and where $\bb{D}^*:=\bb{F}^T\bb{d}^*\bb{F}$, with the eulerian rates $\bb{d}^*=\bb{d}^e$ defined in eqs.(\ref{MultDecAdd})
and (\ref{Decomp}). Then $\bb{D}^{pL} = \bb{D}-\bb{D}^e$ $=\bb{D}-\left(\bb{D}^*+\bb{D}^z\right)$ $=\bb{D}-\left(\bb{D}-\widehat{\bb{D}}^p+\bb{D}^z\right)$ $=\widehat{\bb{D}}^p-\bb{D}^z$;
recalling the relationship in eq.(\ref{DPrefTruesdell2}) and, for simplicity,
employing the stress-strain relationship in eq.(\ref{QuadraticP}), there follows
\begin{equation}
\hspace{-0.5cm}
\begin{array}{l}
\dd\bb{D}^{pL} = \widehat{\bb{D}}^{pL} - \bb{D}^{zpL},\vp\\
\dd\makebox{where}\,\,\,
\bb{D}^{zpL}:=\bb{K}^{-1}:\left(\widetilde{\bb{L}}^p\bb{S}+\bb{S}\widetilde{\bb{L}}^{pT}\right)+
\left(\bb{K}^{-1}:\bb{S}\right)\widetilde{\bb{L}}^p+\widetilde{\bb{L}}^{pT}\left(\bb{K}^{-1}:\bb{S}\right)
\end{array}
\label{ZDPL}
\end{equation}
and where, for a more compact writing, the definition $\widetilde{\bb{L}}^p:=\bb{F}^{p-1}\bb{L}^p\bb{F}^p$ has been employed.

\subsection{A study of the convexity of the activation surface}

\noindent
With slip systems evolving according to eq.(\ref{SlipEvolv}), the activation surface of the crystal is defined
by eq.(\ref{GeneralSchmid2}).
Since slip directions are embedded into the lattice, they are subject to the influence of any deformation history and in particular to that of elastic unloading. A consequence of this is that the elastic domain (at a particle of the crystal)
\emph{is not convex in general}.
To show this, it is will be assumed that the stress-strain relationship is given by that in eq.(\ref{QuadraticP}) and
that non-Schmid effects are absent.
With $\bb{S}$ denoting the current stress state, the current elastic domain of the crystal is defined
as the set of all stress states $\bb{S}^{e}$ that can be reached starting from $\bb{S}$ by a purely
elastic deformation process. Let $\bb{F}$ and $\bb{F}^e$
denote the deformation gradients corresponding to $\bb{S}$ and $\bb{S}^e$, respectively.
With $\bb{S}^e$ corresponding to an elastic state, the plastic state of the crystal at $\bb{S}$ and $\bb{S}^e$ is the same.
Then $\bb{S} = \bb{K}:(\bb{C}-\bb{C}^p)/2$ and
\begin{equation}
\bb{S}^e = \bb{K}:(\bb{C}^e-\bb{C}^p)/2 = \bb{S}+\bb{K}:(\bb{C}^e-\bb{C})/2
\label{A2elastic1}
\end{equation}
Also, it follows that $\bb{F}^e = (\bb{F}^e)^*\bb{F}^p$ and hence
$\bb{C}^e := (\bb{F}^e)^T(\bb{F}^e) = \bb{F}^{pT}(\bb{C}^e)^*\bb{F}^p$.
Then by eq.(\ref{HatDP}) the $\bb{A}$-tensor of slip system $\alpha$ corresponding to the stress state $\bb{S}^e$ is given by the formula
\begin{equation}
\hspace{-3pt}(\bb{A}^{\alpha})^e:=
\frac{1}{2}\bb{F}^{pT}\left[(\bb{C}^e)^*\bb{G}^{\alpha}+\bb{G}^{\alpha T}(\bb{C}^e)^*\right]\bb{F}^p=
\frac{1}{2}\left[\bb{C}^e\bb{H}^{\alpha}+\bb{H}^{\alpha T}\bb{C}^e\right],\,
\label{A2Btensor}
\end{equation}
The elastic domain is then characterized by:
\begin{equation}
\bb{S}^e\cdot\frac{1}{2}\left(\bb{C}^e\bb{H}^{\alpha}+\bb{H}^{\alpha T}\bb{C}^e\right)<\tau^{\alpha}_c,\,
(\forall)\alpha
\label{A2domain}
\end{equation}
Solving eq.(\ref{A2elastic1}) for $\bb{C}^e$ and substituting the result into the above inequality
leads to the following quadratic set of inequalities to be satisfied by $\bb{S}^e$:
\begin{equation}
\phi^{\alpha}(\bb{S}^e)+\bb{S}^e\cdot\left[\left(\bb{C}-2\bb{K}^{-1}:\bb{S}\right)\bb{H}^{\alpha}+\bb{H}^{\alpha T}\left(\bb{C}-2\bb{K}^{-1}:\bb{S}\right)\right]/2<\tau^{\alpha}_c
\end{equation}
where the quadratic term in each inequality reads
\begin{equation}
\phi^{\alpha}(\bb{S}^e):=\left(\bb{K}^{-1}:\bb{S}^e\right)\cdot\left(\bb{S}^e\bb{H}^{\alpha T}+\bb{H}^{\alpha}\bb{S}^e\right)
\label{Aphi1}
\end{equation}
To simplify, let us assume isotropic elastic properties:
$\bb{K}^{-1}=a\bb{I}\otimes\bb{I}+b\overline{\bb{I}}$, with $a := -\lambda/[2\mu(3\lambda+2\mu)]$ and $b := 1/(2\mu)$, in terms
of Lame's parameters. Then
\begin{equation}
\phi^{\alpha}(\bb{S}^e) = \left[a\tr(\bb{S}^e)\bb{S}^e+b(\bb{S}^e)^2\right]\cdot(\bb{H}^{\alpha}+\bb{H}^{\alpha T})
\label{Aphi2}
\end{equation}
The above $\phi^{\alpha}$ is not a convex function. First, let us remark some properties of the $\bb{H}^{\alpha}$ tensor:
\begin{equation}
\bb{H}^{\alpha} = \left(\bb{F}^{p-1}:\bb{m}^{\alpha}\right)\otimes\left(\bb{F}^{pT}:\bb{n}^{\alpha}\right),\,\,
\tr(\bb{H}^{\alpha}) = 0,\,\,\, (\bb{H}^{\alpha}+\bb{H}^{\alpha T})^2\cdot\bb{H^{\alpha}} = 0
\end{equation}
Then, with $\bb{S}_H:=\bb{H}^{\alpha}+\bb{H}^{\alpha T}$, one can further calculate:
\begin{equation}
\phi^{\alpha}(\bb{I}) = 0,\,\,\, \phi^{\alpha}(\bb{S}_H) = 0,\,\,\, \phi^{\alpha}(t_1\bb{I}+t_2\bb{S}_H) = (6a+4b)t_1t_2m^2n^2
\end{equation}
for any reals $t_1$ and $t_2$, where the notations $m:=|\bb{F}^{p-1}:\bb{m}^{\alpha}|$, $n:=|\bb{F}^{pT}:\bb{n}^{\alpha}|$ have been employed.
In particular, with $6a+4b>0$, when $t_1t_2<0$ it follows that $\phi^{\alpha}(t_1\bb{I}+t_2\bb{S}_H)<0$
thus showing that $\phi^{\alpha}$ is not a positive definite quadratic.
More precisely, for each $\alpha$ the nature of the above quadratic is hyperbolic and hence the elastic domain of the crystal,
as defined by eq.(\ref{A2domain}) is not a convex set.

For moderate stresses the quadratic terms gathered in $\phi^{\alpha}$ are barely "visible", their amplitude being several orders
smaller than that of the linear terms. In other words, the overall shape of the elastic domain of the crystal is that resulting
from the linear part of the activation criterion, while the quadratic terms just superimpose small variations upon it.

\subsection{Flow rules and pseudo-deviations form normality}

\noindent
In general, the difference between the two rates in eq.(\ref{ZDPL}) is just a small fraction of the rate of plastic deformation.
However, if non-Schmid effects are to be accounted for, then $\bb{D}^{zpL}$ can be neglected only after a comparison with
the magnitude of these effects. When $\bb{D}^{zpL}$ cannot be neglected,
it is at the origin of an additional source of deviation from the cone of normal directions.
Indeed, by eqs.(\ref{NormalityRef}) and (\ref{ZDPL}) there holds
\begin{equation}
\delta\bb{S}\cdot\left(\bb{D}^{pL}+\bb{D}^{zpL}+\bb{Z}\right)\leq 0
\end{equation}
inequality valid for any elastic direction $\delta\bb{S}$.

Then considering an aggregate of single crystals, averaging the above inequality and employing
eqs.(\ref{MacroDP2}) and (\ref{Moperator2}), results in the inequality
\begin{equation}
\hspace{-1cm}
\delta\overline{\bb{S}}\cdot\left(\overline{\bb{D}}^{pL}+\overline{\bb{D}}^{zpL}+\overline{\bb{Z}}\right)\leq 0,\,\,\,\,
\makebox{with\,\,\,}\,
\overline{\bb{D}}^{zpL}:=
\frac{1}{|\Omega|}\int_{\Omega}\overline{\bb{K}}^{(-1)}\bb{P}^T\bb{K}:\bb{D}^{zpL}\,dX
\label{ZMPFlowrule}
\end{equation}
valid for any macro elastic direction $\delta\overline{\bb{S}}$.
Assuming the macro yield surface is smooth (and convex), the above inequality
then translates into the flow rule
\begin{equation}
\overline{\bb{D}}^{pL}+\overline{\bb{D}}^{zpL}+\overline{\bb{Z}} =
\dot{\lambda}\frac{\partial
f}{\partial\overline{\bb{S}}}\left(\overline{\bb{S}},...\right)
\label{ZZFlowRule}
\end{equation}
with $\dot{\lambda}$ denoting a scalar parameter characterizing
the magnitude of
$\overline{\bb{D}}^{pL}+\overline{\bb{D}}^{zpL}+\overline{\bb{Z}}$.
With $\overline{\bb{D}}^{pL}$ representing the macro-rate of
plastic deformation and $\overline{\bb{Z}}$ the rate of
(intrinsic) non-Schmid effects, the additional term
$\overline{\bb{D}}^{zpL}$ represents a pseudo-deviation from
normality, induced solely by the chosen parametrization of the
plastic state at constituent level.


\end{document}